\documentclass[aps,prd,preprint,superscriptaddress,tightenlines,nofootinbib]{revtex4}

\usepackage{graphicx}
\usepackage{dcolumn}
\usepackage{bm}

\def\Y{\ifmmode \Upsilon \else%
$\Upsilon$%
\fi}
\def\chib{\ifmmode \chi_b \else%
$\chi_b$%
\fi}
\def\chibp{\ifmmode \chi_b' \else%
$\chi_b'$%
\fi}
\def\Q#1#2#3#4{\ifmmode
 \,#1\,{^{#2}#3}_{#4}
\else%
$#1\,{^{#2}#3}_{#4}$ %
\fi}
\def\eonem#1#2{\ifmmode
\left| <#2|r|#1> \right|
\else%
$\left| <#1|r|#2> \right|$
\fi}

\def\ee{\ifmmode e^+e^- \else $e^+e^-$  \fi}
\def\mm{\ifmmode \mu^+\mu^- \else $\mu^+\mu^-$  \fi}
\def\LL{\ifmmode \ell^+\ell^- \else $\ell^+\ell^-$  \fi}

\def\etal{{\it et al.}}
\def\B{{\cal B}}
\def\BR{\B}

\begin{document}

\preprint{CLNS 06/1955}       
\preprint{CLEO 06-04}         

\title{Observation of $\psi(3770)\to\gamma\chi_{c0}$}



\author{R.~A.~Briere}
\author{I.~Brock~\altaffiliation{Current address: Universit\"at Bonn, Nussallee 12, D-53115 Bonn}}
\author{J.~Chen}
\author{T.~Ferguson}
\author{G.~Tatishvili}
\author{H.~Vogel}
\author{M.~E.~Watkins}
\affiliation{Carnegie Mellon University, Pittsburgh, Pennsylvania 15213}
\author{J.~L.~Rosner}
\affiliation{Enrico Fermi Institute, University of
Chicago, Chicago, Illinois 60637}
% ...
\collaboration{CLEO Collaboration} 
\noaffiliation


\date{\today}

\begin{abstract}
From $\ee$ collision data acquired with the CLEO-c detector at CESR,
we search for the non-$D\bar D$ decays $\psi(3770)\to\gamma\chi_{cJ}$,
with $\chi_{cJ}$ reconstructed in four 
exclusive decays modes containing charged pions and kaons.
We report the first observation of such decays for $J=0$
with a branching ratio of $(0.73\pm0.07\pm0.06)\%$.
The rates for different $J$ are consistent with 
the expectations assuming $\psi(3770)$ is predominantly a $1^3D_1$ state
of charmonium, but only if relativistic corrections are applied.
\end{abstract}

\maketitle

Observation of the narrow $X(3872)$ and $Y(4260)$ states \cite{XY}
above open charm threshold, and their possible interpretation as
states beyond the traditional $c\bar c$ model of 
charmonium \cite{charmonium},
calls for thorough investigation of the 
lightest charmonium state above the $D\bar D$ threshold - $\psi(3770)$.
The common interpretation of the $\psi(3770)$ assumes it is
predominantly the $1^3D_1$ $c\bar c$ state, with a small admixture 
of $2^3S_1$. Except for the large $D\bar D$ decay width and 
rough agreement with the potential model mass predictions, 
there have been no other experimental data to verify this assumption.
Although decays of $\psi(3770)$ to $\pi^+\pi^- J/\psi$, $\pi^0\pi^0 J/\psi$ 
and $\eta J/\psi$ have been measured to be non-zero \cite{BES,gammaee}, 
such hadronic modes present a less sensitive probe of the 
charmonium model than rates for 
$\psi(3770)\to\gamma\chi_{cJ}$ since they involve 
hadronization probabilities.

Previously, we have reported observation of 
$\psi(3770)\to\gamma\chi_{c1}$ with $\chi_{c1}\to\gamma J/\psi$, 
$J/\psi\to l^+l^-$ \cite{psipp2gX1}.
The branching ratio for
$\psi(3770)\to\gamma\chi_{c0}$ is predicted to be the 
largest \cite{Chao,Rosner,Eichten,Barnes}, but the small branching ratio for
$\chi_{c0}\to \gamma J/\psi$ reduces the sensitivity so much that
only a loose upper limit could be set in Ref.~\cite{psipp2gX1}.
However, hadronic $\chi_{c0}$ decays are copious and thereby offer
complementary probes for these photon transitions.
Backgrounds from $D\bar D$ decays and continuum processes are suppressed 
by full reconstruction of $\chi_{cJ}$ decays to
a few exclusive hadronic final states.
We use the following decay modes:
$\chi_{cJ}\to K^+ K^-$ $(2K)$,
$\chi_{cJ}\to \pi^+\pi^-\pi^+\pi^-$ $(4\pi)$,
$\chi_{cJ}\to K^+ K^-\pi^+\pi^-$ $(2K2\pi)$ and
$\chi_{cJ}\to \pi^+\pi^-\pi^+\pi^-\pi^+\pi^-$ $(6\pi)$.
To minimize sensitivity to large uncertainties in branching fractions and
resonant substructure for these channels, we measure the rates relative to 
those seen in $\psi(2S)$ decays with the same detector,
$$
R_J \equiv
\frac{\BR(\psi(3770)\to\gamma\chi_{cJ})\times\BR(\chi_{cJ}\to\pi^{\pm},K^{\pm})}{
\BR(\psi(2S)\to\gamma\chi_{cJ})\times\BR(\chi_{cJ}\to\pi^{\pm},K^{\pm})},
$$
and normalize to 
${\cal B}(\psi(2S)\to\gamma\chi_{cJ})$ \cite{psip2gXj},
which was measured by fitting 
inclusive photon energy spectra. 
Thus, our results for $\BR(\psi(3770)\to\gamma\chi_{cJ})$ 
are not only
independent of $\BR(\chi_{cJ}\to\pi^{\pm},K^{\pm})$, but also 
depend only on ratios of detection efficiencies for $\psi(3770)$ 
and $\psi(2S)$. The latter are almost independent of the resonant 
substructure and, therefore, can be more reliably determined.

The data were acquired at a center-of-mass energy of 3773 MeV with
the CLEO-c detector \cite{CLEOdet} operating at the Cornell Electron
Storage Ring (CESR), and correspond to an integrated luminosity 
(number of resonant decays) of 281 pb$^{-1}$ ($(1.80\pm0.05)\times10^6$) 
at the $\psi(3770)$ and 2.9 pb$^{-1}$ ($(1.51\pm0.05)\times10^6$) 
at the $\psi(2S)$.
The CLEO-c detector features a solid angle coverage of 93\%\ for charged
and neutral particles. The cesium iodide (CsI) calorimeter attains
photon energy resolutions of 2.2\%\ at $E_\gamma=1$ GeV and 5\%\ at
100 MeV. For the data presented here, the charged particle tracking
system operates in a 1.0 T magnetic field along the beam axis and
achieves a momentum resolution of 0.6\%\ at p = 1 GeV.
Particle identification is performed using Ring-Imaging Cherenkov 
Detector (RICH) in combination with specific ionization loss (dE/dx) in 
the gaseous tracking volume.

\begin{table}
\caption{Efficiencies for $\psi(2S)/\psi(3770)\to\gamma\chi_{cJ}$,
$\chi_{cJ}\to\pi^{\pm}, K^{\pm}$, based on Monte Carlo of phase-space
$\chi_{cJ}$ decays (i.e.\ no intermediate resonances).
 \label{tab:eff}}
\begin{center}
\begin{tabular}{rc|ccc}
\hline\hline
               &       & \multicolumn{3}{c}{Efficiency (\%) }\\
               &      &$J=2$&$J=1$&$J=0$ \\
\hline
           $\psi(2S)\to\gamma\chi_{cJ}\to$
               &$4\pi$  &$33$ &$35$ &$34$\\
               &$2K2\pi$&$25$ &$27$ &$28$\\
               &$6\pi$  &$23$ &$25$ &$27$\\
               &$2K$    &$43$ &$44$ &$42$\\
      $\psi(3770)\to\gamma\chi_{cJ}\to$
               &$4\pi$  &$35$ &$36$ &$34$\\
               &$2K2\pi$&$29$ &$30$ &$29$\\
               &$6\pi$  &$27$ &$28$ &$27$\\
               &$2K$    &$44$ &$44$ &$41$\\
\hline\hline
\end{tabular}
\end{center}
\vskip-0.5cm
\end{table}

We select events with exactly 6, 4 or 2 charged
tracks and at least one photon candidate with energy
above 60 MeV.
The highest energy photon is considered to be the signal photon,
while other neutral clusters in the calorimeter 
are considered fragments of hadronic
showers, and therefore ignored. We separate pions and kaons using
a log-likelihood difference, which optimally combines the dE/dX and RICH
information. The track is considered a kaon if the kaon hypothesis
is more likely. The RICH information is used only if the track
momentum is above kaon radiation threshold (700 MeV) and the number of
Cherenkov photons for the kaon hypothesis is required to be at least 3
in this case. We also impose 3$\sigma$ consistency on dE/dx. Those
tracks not identified as kaons become pion candidates if they
satisfy 3$\sigma$ consistency with dE/dX. Events with odd numbers
of kaons or pions are rejected. The total energy and Cartesian
components of momentum of the selected charged particles and the
photon must be consistent within $\pm30$ MeV with the expected
center-of-mass four-vector components, which take into account a small beam
crossing angle. To improve resolution on the photon energy,
we then constrain these quantities to the expected
values via kinematic fitting of events.
Selection efficiencies obtained with {\sc GEANT} \cite{GEANT}
based simulation of detector response are given in Table~\ref{tab:eff}.

\begin{figure}[htbp]
\includegraphics[width=\hsize]{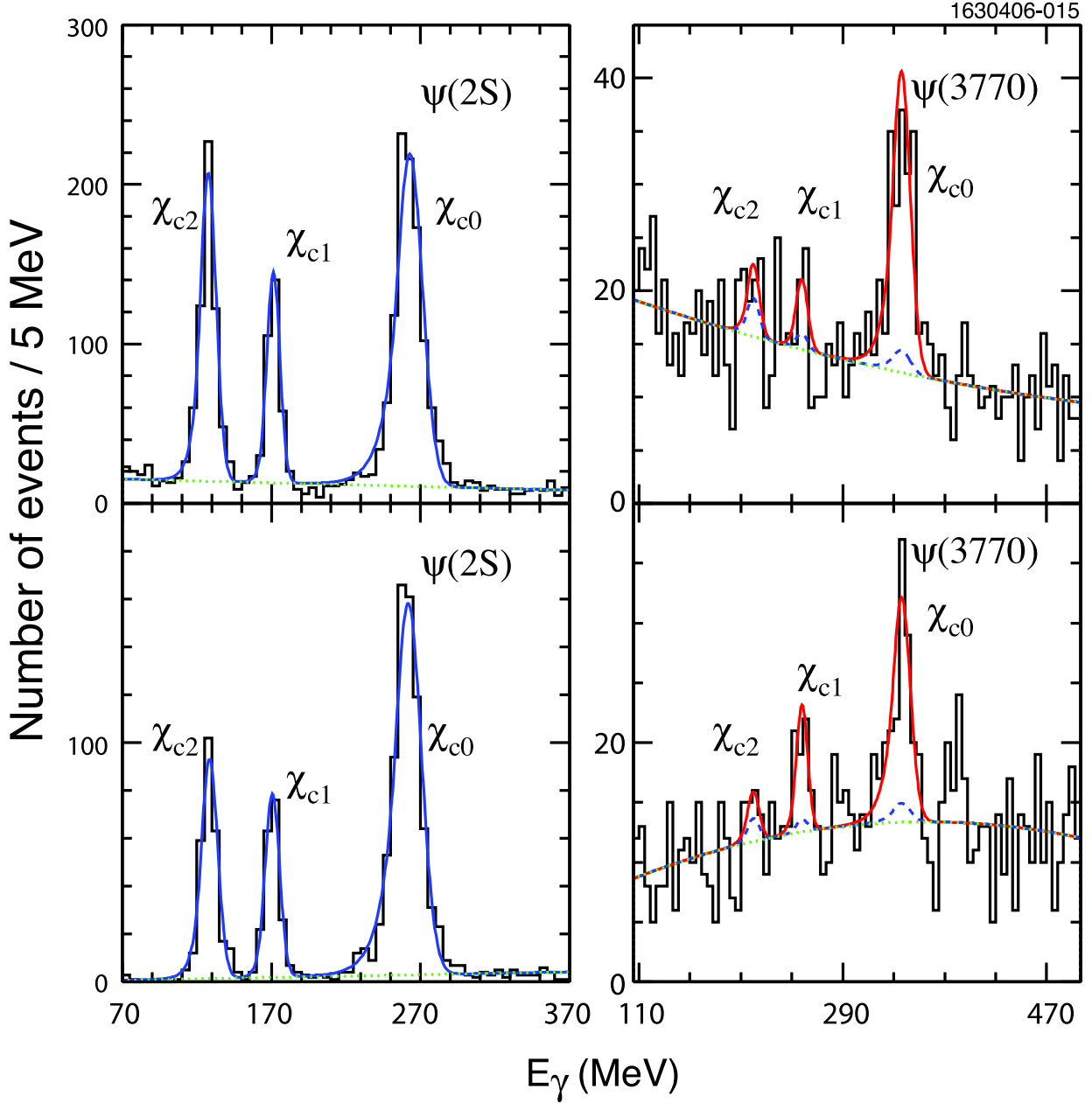} 
\caption{
Distribution of photon energy for $4\pi$ (top) and
$2K2\pi$ (bottom) decay samples in CLEO-c $\psi(2S)$ (left) and
$\psi(3770)$ (right) data. 
Solid histogram is data, 
smooth curve is fit to the data.
Dashed line shows radiative return
background contribution from $\psi(2S)$ tail and 
dotted line is polynomial background.
\label{fig:pe12} }
\end{figure}

\begin{figure}[htbp]
\includegraphics[width=\hsize]{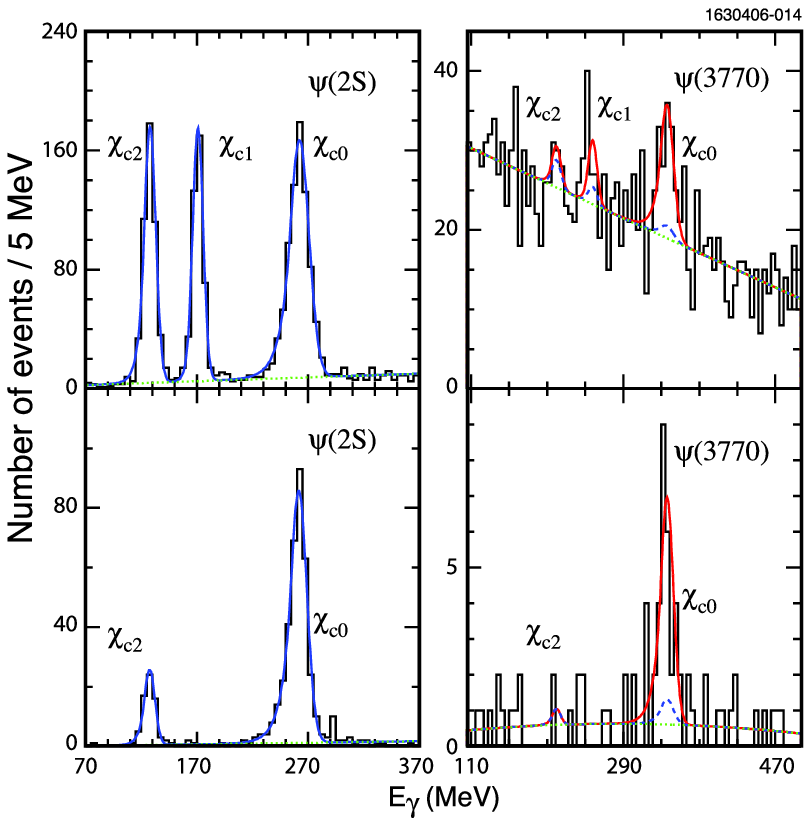} 
\caption{Distribution of photon energy for $6\pi$ (top) and $2K$
(bottom) decay samples in CLEO-c $\psi(2S)$ (left) and $\psi(3770)$
(right) data. 
Solid histogram is data, 
smooth curve is fit to the data.
Dashed line shows radiative return
background contribution from $\psi(2S)$ tail and 
dotted line is polynomial background.
\label{fig:pe34} }
\end{figure}

The energy of the photon candidates is plotted for the data for different
decay channels in  Fig.~\ref{fig:pe12} and
Fig.~\ref{fig:pe34}. Fits used to extract signal amplitudes are
also shown.
Each photon line is represented by a detector
response function, parameterized by the so-called Crystal Ball line
(CBL) shape. CBL is a Gaussian (described by the peak energy,
$E_{0}$, and energy resolution, $\sigma_{E}$) turning into a power
law tail, $1/(E_{0}-E+\text{const})^n$, at an energy of
$E_{0}-\alpha\sigma_{E}$.  We fix
$\alpha$ and $n$ to the values  determined from the signal Monte Carlo.
The peak amplitude ($A_{\psi(2S)}^{\text{in}~\psi(2S)}$), 
peak energy and widths are free parameters in
the fit to the $\psi(2S)$ data. 
The smooth background is represented by a first order polynomial. 
In the fit to the $\psi(3770)$ data only the 
peak amplitudes ($A_{\psi(3770)}$) are free
parameters, while the CBL parameters are fixed to the
predictions from the signal Monte Carlo.
In addition to the smooth backgrounds, represented by a second order
polynomial, the $\psi(3770)$ data
also contain radiatively produced $\psi(2S)$ background.
After our selection cuts, the latter cannot be distinguished from the
$\psi(3770)$ signal. They are explicitly represented in the fit by peaks with
the amplitudes, $A_{\psi(2S)}^{\text{in}~{\psi(3770)}}$, 
fixed to the values estimated from the $\psi(2S)$ 
data ($A_{\psi(2S)}^{\text{in}~{\psi(2S)}}$) and
extrapolated to the $\psi(3770)$ beam energy with help 
of the theoretical formulae:
$$ A_{\psi(2S)}^{\text{in}~{\psi(3770)}} = {\cal L}_{\psi(3770)} \cdot \epsilon_{\psi(3770)} \cdot
   {\cal B}_X \cdot \Gamma_{ee}(\psi(2S)) \cdot I(s) $$
$$ I(s) = \int_0^{x_{cut}}
   W(s,x) \cdot b(s'(x)) \cdot F_X(s'(x)) dx. $$
Here, we are using the same notation as in Ref.~\cite{gammaee}:
${\cal L}$ is the integrated luminosity; 
$\epsilon$ is the efficiency; 
${\cal B}_X$  is the branching ratio for
$\psi(2S)\to\gamma\chi_{cJ}\to\gamma X$ ($X$ is the hadronic final
state) at the $\psi(2S)$ resonance peak; $x$ is energy radiated in
$e^+e^-\to \gamma\psi(2S)$ divided by its maximal possible value
(i.e. by $E_{\text{beam}}=\sqrt{s}/2$); $s'$ is the mass-squared with which
the $\psi(2S)$ is produced ($s'(x)=s(1-x)$); $W(s,x)$ is the initial
state radiation probability (see Ref.~\cite{gammaee}\ for the
definition and discussion); $b(s')$ is the relativistic Breit-Wigner
formula describing the $\psi(2S)$ resonance
($b(s')=12\pi\Gamma_R/[(s'-M^2_R)^2+M^2_R \Gamma^2_R]$); and
$F_X(s')$ is the phase-space factor between the $\psi(2S)$ produced
with $\sqrt{s'}$ mass and with its nominal mass, $M_R$. $F_X(s')$
is equal \cite{E1} to 
$(E_\gamma(s')/E_\gamma(M_R^2))^3$, where $E_\gamma$ is
the photon energy in $\psi(2S)\to\gamma\chi_{cJ}$ decay. The
$\psi(2S)$ nominal mass ($M_R$) and total width ($\Gamma_R$) are
taken from PDG \cite{PDG}, 
while $\Gamma_{ee}(\psi(2S))$ is taken from the CLEO
determination utilizing $e^+e^-\to\gamma\psi(2S)$ at 
$E_{CM}=3773$ MeV with
$\psi(2S)$ decaying to $J/\psi$ through a hadronic 
transition \cite{gammaee}. The radiative flux, $W(s,x)$, strongly peaks for
$x\to0$ making the $\psi(2S)$ background indistinguishable from
the $\psi(3770)$ signal within our photon energy resolution.
Unlike in our $X=\gamma J/\psi$ analysis \cite{psipp2gX1}, 
where we used the published CLEO results for
${\cal B}_X$ and relied on the absolute value of the detection efficiency
($\epsilon_\psi(3770)$), in this analysis we set
$$
{\cal B}_X= \frac{A_{\psi(2S)}^{\text{in}~\psi(2S)}}{\epsilon_{\psi(2S)}\cdot
N_{\psi(2S)}},
$$
where $A_{\psi(2S)}^{\text{in}~\psi(2S)}$ is the signal
yield in the fit to the $\psi(2S)$ data. Therefore, our estimates of
the $\psi(2S)$ radiative tail background,
$$ A_{\psi(2S)}^{\text{in}~\psi(3770)} = A_{\psi(2S)}^{\text{in}~\psi(2S)} \cdot
   \frac{\epsilon_{\psi(3770)}}{\epsilon_{\psi(2S)}} \cdot
   \frac{{\cal L}_{\psi(3770)}}{N_{\psi(2S)}} \cdot
   \Gamma_{ee}(\psi(2S)) \cdot I(s), $$
do not rely on absolute values of efficiencies, but only on their
ratio between the $\psi(3770)$ and $\psi(2S)$ data samples. The
upper range of integration in the definition of $I(s)$ is
$x_{\text{cut}}\approx 30$ MeV/$1887$ MeV=$0.016$, because of our cuts on
total energy and momentum.
The signal yields in the $\psi(2S)$ and $\psi(3770)$ data 
are given in
Table~\ref{tab:data}.

\begin{table}
\caption{Fitted signal yields for 
$\psi(2S)/\psi(3770)\to\gamma\chi_{cJ}$,
$\chi_{cJ}\to\pi^{\pm}, K^{\pm}$. The total number of the estimated 
$\psi(2S)$ background events in the $\psi(3770)$ data 
($A_{\psi(2S)}^{\text{in}~{\psi(3770)}}$) is also given.
The errors on the latter quantities are systematic.
All other errors are statistical.
\label{tab:data}}
\begin{center}
\begin{tabular}{r|cccc}
\hline\hline
               &Decay   & \multicolumn{3}{c}{Events}\\
               &Mode    &$J=2$      &$J=1$      &$J=0$ \\

\hline
               &$4\pi$  &$534\pm27$ &$291\pm19$ &$981\pm36$\\
               &$2K2\pi$&$261\pm16$ &$187\pm14$ &$745\pm29$\\
$A_{\psi(2S)}^{\text{in}~\psi(2S)}$
               &$6\pi$  &$469\pm23$ &$408\pm21$ &$744\pm30$\\
               &$2K$    &$64\pm8$   &$-$        &$346\pm19$\\
               & All   &$1329\pm40$&$886\pm32$ &$2816\pm58$\\
 & & & & \\
$A_{\psi(2S)}^{\text{in}~\psi(3770)}$
               & All   &$25\pm6$       &$12\pm3$       &$25\pm6$\\
 & & & & \\
               &$4\pi$  &$9\pm10$   &$14\pm9$   &$112\pm16$\\
               &$2K2\pi$&$6\pm8$    &$25\pm9$   &$73\pm14$\\
$A_{\psi(3770)}$
               &$6\pi$  &$5\pm12$   &$16\pm11$  &$65\pm16$\\
               &$2K$    &$0\pm1$    &$-$        &$24\pm6$\\
               & All   &$20\pm18$  &$54\pm17$  &$274\pm27$\\
\hline\hline
\end{tabular}
\end{center}
\vskip-0.5cm
\end{table}

\begin{figure}[htbp]
\includegraphics[width=\hsize]{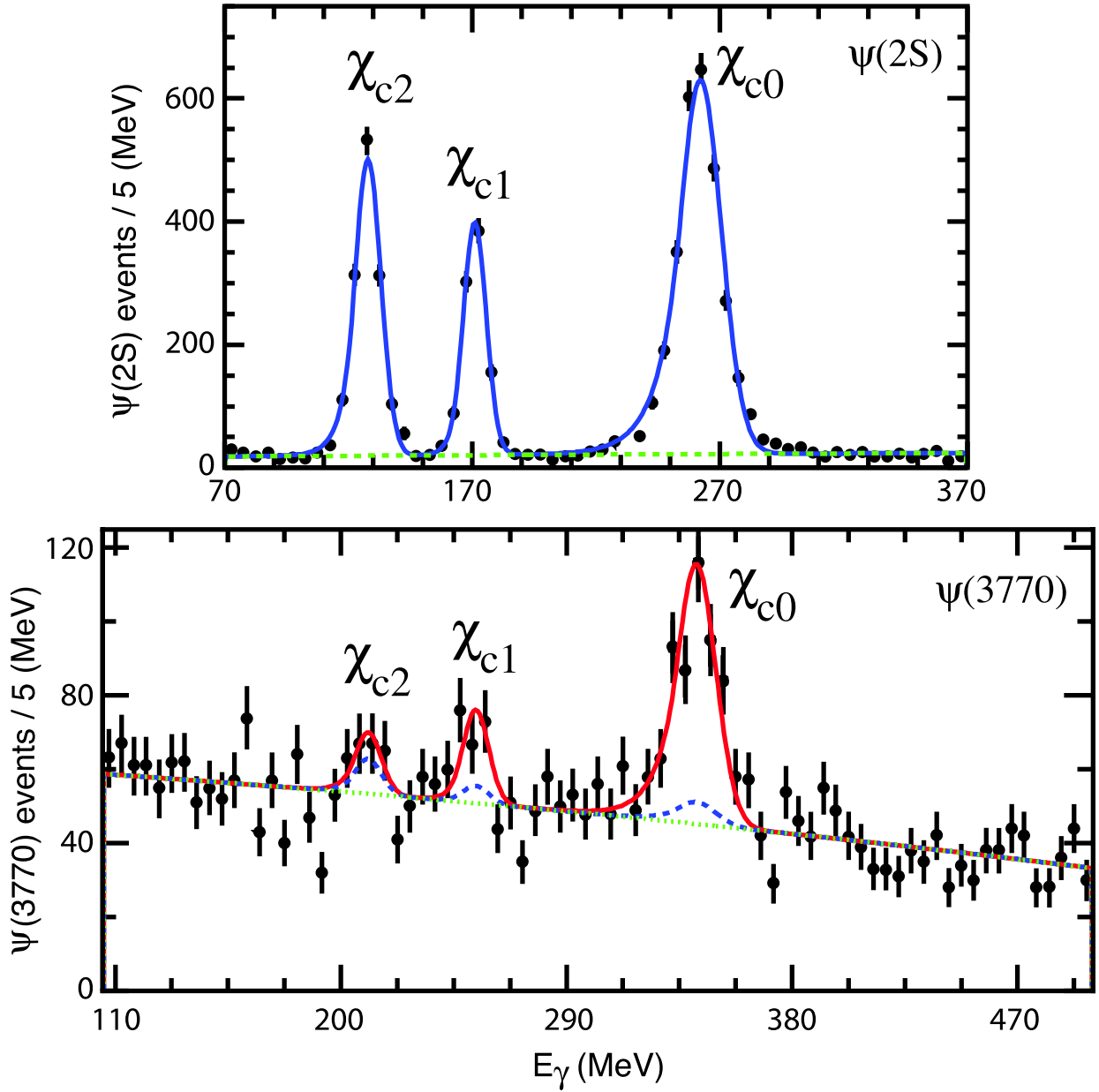} 
\caption{Distribution of photon energy in CLEO-c $\psi(2S)$ (top)
and $\psi(3770)$ (bottom) data summed over all analyzed modes (data
points). The smooth curve shows the sum of the fits performed to the
individual modes. The dashed curve shows the radiative tail from
$\psi(2S)$. The dotted line shows the polynomial background.
\label{fig:pe1234} }
\end{figure}

The results for the ratio of branching ratios, $R_J$,
for individual decay modes
are given in Table~\ref{tab:ratio}. Average values are calculated
using inverse-of-statistical-errors-squared for weights. To estimate
the statistical significance of $\psi(3770)\to\gamma\chi_{cJ}$ signals,
we fit the $\psi(3770)$ data with the background contribution alone
and compare the fit likelihoods to our nominal fits. Combining
likelihoods for all the channels, we obtain statistical significance
of $1.3, 3.6$ and $12.6$ standard deviations for $J=2$, $1$ and $0$,
respectively. The sum of the photon spectra over the individual channels
is shown for $\psi(2S)$ and $\psi(3770)$ data in
Fig.~\ref{fig:pe1234}. Since no significant signal is observed for
$J=2$, we set an upper limit for this state.

\begin{table}
\caption{The ratio
$R_J=\BR(\psi(3770)\to\gamma\chi_{cJ},\chi_{cJ}\to\pi^{\pm},K^{\pm})/
\BR(\psi(2S)\to\gamma\chi_{cJ},\chi_{cJ}\to\pi^{\pm},K^{\pm})$. Only
statistical errors are given here.
 \label{tab:ratio}}
\def\2#1{{\scriptsize #1}}
\begin{center}
\begin{tabular}{c|ccc}
\hline\hline
               Decay   & \multicolumn{3}{c}{$R_J$ in \%}\\
               mode    &$J=2$            &$J=1$         &$J=0$ \\
\hline
               $4\pi$   &$1.3\pm1.5$    &$3.8\pm2.6$    &$9.6\pm1.4$\\
               $2K2\pi$ &$1.7\pm2.4$    &$9.9\pm4.0$    &$8.2\pm1.7$\\
               $6\pi$   &$0.7\pm1.8$    &$2.9\pm2.2$    &$7.4\pm1.8$\\
               $2K$     &$0.0\pm1.4$    &$-$            &$6.0\pm1.6$\\
  Average               &$0.8\pm0.8$    &$4.3\pm1.6$    &$7.9\pm0.8$\\
\hline\hline
\end{tabular}
\end{center}
\vskip-0.5cm
\end{table}

\begin{table}
\caption{Systematic errors and their sources. \label{tab:sys}}
\def\2#1{{\scriptsize #1}}
\begin{center}
\def\1#1{\multicolumn{1}{r|}{#1}}
\begin{tabular}{l|ccc}
\hline\hline
                     & \multicolumn{3}{c}{Relative change in \%}\\
                            &$J=2$ &$J=1$ &$J=0$\\
\hline
 Luminosity                   &$1$   &$1$   &$1$\\
 $\psi(3770)$ cross-section   &$3$   &$3$   &$3$\\
 Number of $\psi(2S)$ decays  &$3$   &$3$   &$3$\\
 Resonant substructure        &$2$   &$<1$  &$<1$\\
 $\pm25\%$ change in $\psi(2S)$ bkg.  &$39$  &$6$   &$2$\\
 & & & \\
\multicolumn{1}{r|}{Fit systematics} & & & \\
 $\pm7\%$ change in $\sigma_E$ &$10$  &$8$   &$4$\\
 $\pm10\%$ change in fit range &$17$  &$5$   &$1$\\
 Using Gaussian signal shape   &$9$   &$2$   &$1$\\
 Decreasing bin-size to half   &$15$  &$3$   &$<1$\\
 $\pm1$ order of bkg.\ polynomial&$47$  &$9$   &$2$\\
 Total fit systematics         &$53$  &$12$  &$5$\\
 & & & \\
{Total systematic error on $R_J$}       &$66$  &$14$  &$7$\\
 & & & \\
 $\BR(\psi(2S)\to\gamma\chi_{cJ})$ &$6$   &$5$   &$4$\\
 Number of $\psi(2S)$ decays  &$-3$  &$-3$  &$-3$\\
{Total systematic error on}         &&&\\
\multicolumn{1}{r|}{${\BR(\psi(3770)\to\gamma\chi_{cJ})}$}
&${66}$ &${15}$  &${8}$\\
\hline\hline
\end{tabular}
\end{center}
\vskip-0.5cm
\end{table}

Various contributions to the systematic errors are
listed in Table~\ref{tab:sys}. 
We simulated signal events assuming various
resonant substructures and compared  
the efficiency ratio to our nominal values 
obtained with the phase-space model
to evaluate the error in efficiency simulation. 
Including the systematic errors, our results for
the ratio of branching ratios are: 
$R_0=(7.9\pm0.8\pm0.6)\%$, 
$R_1=(4.3\pm1.6\pm0.6)\%$ and
$R_2<2.2\%$ (90\%\ C.L.).  
The 3\%\ uncertainty in the number of $\psi(2S)$
resonant decays contributes to the $R_J$ measurement, but 
cancels when multiplied by the
inclusively measured $\BR(\psi(2S)\to\gamma\chi_{cJ})$ \cite{psip2gXj}.
The results for $\BR(\psi(3770)\to\gamma\chi_{cJ})$
are  $(0.73\pm0.07\pm0.06)\%$, $(0.39\pm0.14\pm0.06)\%$
and  $<0.20\%$  (90\% C.L.) for $J=0,1$ and $2$, respectively.
They are consistent with the results obtained previously 
by CLEO \cite{psipp2gX1} 
using $\chi_{cJ}\to\gamma J/\psi$ decays:
$<4.4\%$ (90\%\ C.L.), $(0.28\pm0.05\pm0.04)\%$ and 
$<0.09\%$  (90\% C.L.), correspondingly.
The two analyses are complementary. 
While this analysis offers much better sensitivity for $J=0$,
the previous analysis is more sensitive for $J=1$ and $2$.
The $J=1$ signal is observed in both analyses.
Combining both analyses we obtain 
$\BR(\psi(3770)\to\gamma\chi_{c1})=(0.29\pm0.05\pm0.04)\%$.

We turn the branching ratio results to transition widths using
$\Gamma_{\text{tot}}=(23.6\pm2.7)$ MeV from PDG \cite{PDG}. 
The results are given in Table~\ref{tab:widths}, where they are
compared to theoretical predictions.

\begin{table}
\caption{Our measurements
of the photon transitions widths
(statistical and systematic errors)
compared to theoretical predictions.
The $J\!=\!0$ measurement comes from this analysis.
The $J\!=\!2$ upper limit comes from Ref.\cite{psipp2gX1}.
The $J\!=\!1$ measurement comes from the combination of this analysis and
of the result in Ref.\cite{psipp2gX1}.
 \label{tab:widths}}
\begin{center}
\def\1#1{\multicolumn{1}{r|}{#1}}
\begin{tabular}{l|ccc}
\hline\hline
  & \multicolumn{3}{c}{$\Gamma(\psi(3770)\to\gamma\chi_{cJ})$ in keV}\\
                                    &$J=2$     &$J=1$      &$J=0$\\
\hline
Our results &$<21$ & $70\pm17$ & $172\pm30$ \\
& & & \\
Rosner (non-relativistic) \cite{Rosner}         &$24\pm4$  &$73\pm9$   &$523\pm12$ \\
Ding-Qin-Chao \cite{Chao}  &&&\\
\1{non-relativistic}         &$3.6$     &$95$      &$312$ \\
\1{relativistic}             &$3.0$     &$72$       &$199$ \\
Eichten-Lane-Quigg \cite{Eichten}    &&&\\
\1{non-relativistic}                              &$3.2$     &$183$      &$254$ \\
\1{with coupled-channels corrections}  &$3.9$     &$59$       &$225$ \\
Barnes-Godfrey-Swanson \cite{Barnes} &&&\\
\1{non-relativistic}         &$4.9$     &$125$      &$403$ \\
\1{relativistic}             &$3.3$     &$77$       &$213$ \\
\hline\hline
\end{tabular}
\end{center}
\vskip-0.5cm
\end{table}

\def\mat{<\!1^3P_J |r| 1^3D_1\!>}
The theoretical predictions are based on 
potential model calculations \cite{E1}
of the electric dipole matrix element $\mat$:
$$ \Gamma_{J} = \frac{4}{3} e_Q^2 \alpha E_\gamma^3 C_J \mat^2, $$
where $e_Q$ is the $c$ quark charge and $\alpha$ is the fine structure
constant. The spin factors $C_J$ are equal to 
$2/9$, $1/6$ and $1/90$ for $J=0$, $1$ and $2$, 
respectively \cite{KwongRosner}. 
The phase-space factor ($E_\gamma^3$) also favors the $J=0$
transition. Together, the spin and phase-space factors predict
enhancement of the $J=0$ width 
by a factor of $\sim3.2$ and $\sim85$ over $J=1$ and
$J=2$, respectively. 
In the non-relativistic limit, the matrix element is
independent of $J$. 
The measured ratios of the widths,
$\Gamma_0/\Gamma_1=2.5\pm0.6$ and $\Gamma_0/\Gamma_2>8$ (90\%\ C.L.),
are consistent with these crude predictions,
therefore, providing further evidence that $\psi(3770)$ is 
predominantly a $1^3D_1$ state. 
A small admixture of $2^3S_1$ wave, necessary to explain the observed
$\Gamma_{ee}(\psi(3770))$, 
is expected to increase $\Gamma_0$ and $\Gamma_2$ while making
$\Gamma_1$ smaller \cite{Chao,Rosner}.
The large experimental and theoretical uncertainties in 
$\Gamma_J$ make testing of
the mixing hypothesis via radiative transitions
difficult.

As evident from Table~\ref{tab:widths}, the naive
non-relativistic calculations tend to overestimate 
absolute values of the transition rates.
Relativistic \cite{Chao,Barnes} or coupled-channel \cite{Eichten}
corrections are necessary for quantitative agreement with the
data.
The latter is not surprising since non-relativistic calculations
also overestimate $\psi(2S)\to\gamma\chi_{cJ}$ 
transition rates \cite{Skwarnicki}.

We gratefully acknowledge the effort of the CESR staff 
in providing us with excellent luminosity and running conditions. 
This work was supported by 
the A.P.~Sloan Foundation,
the National Science Foundation,
the U.S. Department of Energy, and
the Natural Sciences and Engineering Research Council of Canada.

\end{document}